\documentclass[wssquare]{ws-procs961x669} 
\usepackage{ws-toc,ws-multind}
\usepackage{aas_macros}
\usepackage{placeins}

\usepackage{hyperref} 
\hypersetup{
 colorlinks=true,
 citecolor=blue,
 linkcolor=blue,
 urlcolor=blue,
 }

\makeindex{author} 

\begin{document}

\wstoc{Leveraging Transfer Learning for Astronomical Image Analysis}{S. Cavuoti}

\title{Leveraging Transfer Learning for Astronomical Image Analysis}

\author{Stefano Cavuoti}
\address{INAF – Astronomical Observatory of Capodimonte\\ 
Salita Moiariello 16, I-80131 Napoli, Italy\\
\email{stefano.cavuoti@inaf.it}}

\author{Lars Doorenbos}
\address{AIMI, ARTORG Center, University of Bern, \\Murtenstrasse 50, CH-3008 Bern, Switzerland}

\author{Demetra De Cicco, Gianluca Sasanelli, Massimo Brescia, Giuseppe Longo, Maurizio Paolillo}
\address{Department of Physics ‘Ettore Pancini’, University Federico II\\ Via Cintia, 21, I-80126 Napoli, Italy\\
}

\author{Olena Torbaniuk }
\address{ Department of Physics and Astronomy `Augusto Righi', University of Bologna, Via Gobetti 93/2, I-40129 Bologna, Italy\\}

\author{Giuseppe Angora, Crescenzo Tortora}
\address{INAF – Astronomical Observatory of Capodimonte\\ 
Salita Moiariello 16, I-80131 Napoli, Italy\\}

\begin{abstract}
The exponential growth of astronomical data from large-scale surveys has created both opportunities and challenges for the astrophysics community. This paper explores the possibilities offered by transfer learning techniques in addressing these challenges across various domains of astronomical research. We present a set of recent applications of transfer learning methods  for astronomical tasks based on the usage of a pre-trained convolutional neural networks. The examples shortly discussed include the detection of candidate active galactic nuclei (AGN), the possibility of deriving physical parameters for galaxies directly from images, the identification of artifacts in time series images, and the detection of strong lensing candidates and outliers. We demonstrate how transfer learning enables efficient analysis of complex astronomical phenomena, particularly in scenarios where labeled data is scarce. This kind of method will be very helpful for upcoming large-scale surveys like the Rubin Legacy Survey of Space and Time (LSST). By showcasing successful implementations and discussing methodological approaches, we highlight the versatility and effectiveness of such techniques.
\end{abstract}

\bodymatter

\section{Introduction}\label{intro}

In recent years, the field of astronomy has witnessed an unprecedented growth in the volume and complexity of observational data. Wide-field surveys such as the Sloan Digital Sky Survey (SDSS, \cite{York2000}), the Kilo Degree Square Survey (KiDS, \cite{deJong2015}), and the Hyper Suprime-Cam Subaru Strategic Program (HSC SSP, \cite{Aihara2019}) on the one hand have dramatically expanded our understanding of the universe, on the other hand, they also increased the average size of the dataset involved in astronomical analysis. The next years with the new catalogues coming from projects like the Vera C. Rubin Observatory Legacy Survey of Space and Time (Rubin-LSST, \cite{Ivezic2019}), Euclid \cite{scaramella2022euclid}, and the James Webb Space Telescope (JWST, \cite{AlvarezMarquez2019}) promise to further increase the scale of astronomical data by orders of magnitude.

This data deluge presents, of course, both opportunities and challenges. While it offers the potential for groundbreaking discoveries, it also necessitates the development of sophisticated analytical techniques capable of efficiently processing and interpreting vast amounts of information. Traditional methods of data analysis can become inadequate in the face of this ``big data'' tsunami, leading to a surge of interest in machine learning (ML) and deep learning (DL) approaches in astronomy \cite{Baron2019, Longo2019, Fluke2020}.

Transfer learning, among the various paradigms, allows the leveraging of knowledge gained from one domain or task to improve performance on a different domain or task where there is some sort of similarity. This approach can be precious in astronomy, where labeled data for some specific use cases can be scarce despite the abundance of unlabeled data. For instance, the task of classifying objects from images is really very common, and this is something that is done in several different domains.


In this paper, we recap some of our transfer learning applications in astronomy, demonstrating how this approach can be used to tackle a wide range of challenges, from object detection and classification to artifact removal in time series data and source characterization. 

\section{The model}

At the core of our transfer learning approach in astronomy lies the feature extraction that we presented at first in the paper \cite{Doorenbos22} where the tool \texttt{ULISSE} (aUtomatic Lightweight Intelligent System for Sky Exploration), a versatile framework for feature extraction and similarity search, has been introduced. The key components and functionality of our approach can be summarized as follows:

\subsection{Core Principles}

\begin{enumerate}
 \item \textbf{Feature Extraction:} our methods employ \texttt{EfficientNet} \cite{tan2019efficientnet}, a convolutional neural network\footnote{Convolutional neural networks typically consist of two parts \cite{Schmidhuber2015}: the first part transforms the input image into a feature vector through a series of convolutional layers, pooling layers, and activation functions, effectively functioning as a feature extractor. The second part takes these features and uses them to perform the actual classification task, usually through a multi-layer perceptron (MLP, \cite{McCulloch1943,Rosenblatt1958} neural network).

} pre-trained on the \texttt{ImageNet}\footnote{\texttt{ImageNet} is a large-scale dataset which contains around 1.3 million images with 1000 classes. \url{https://www.image-net.org/}} \cite{deng2009imagenet} dataset, we eliminate the classificator part and consider only the first part of the network that becomes a feature extractor. This network transforms input astronomical images into a high-dimensional feature space (1280 features for each image).
 \item \textbf{Feature Space:} The extracted features create a new parameter space where astronomical objects can be represented as points. This transformation allows for the definition and computation of distances between objects in this feature space.
 \item \textbf{Similarity/Dissimilarity Search:} In this feature space, the similarity between astronomical objects can be quantified by measuring the distance between their corresponding points. In this way, it is possible to define the concept of most similar/dissimilar that will tackle the specific problem to be solved
\end{enumerate}

\subsection{Advantages}

This approach offers several key advantages:

\begin{enumerate}
 \item \textbf{Transfer Learning:} By using a pre-trained network, this approach leverages knowledge from a large dataset (\texttt{ImageNet}) and applies it to astronomical data, overcoming the limitation of small labeled datasets since there is no need for any labeled dataset in order to train a complex model; labels can be used just to assess the results.
 \item \textbf{Flexibility:} While the core feature extraction remains consistent, the downstream tasks (e.g., similarity search, anomaly detection, classification) can be tailored to specific problems.
\end{enumerate}

In essence, with this approach, we transform the problem of analyzing a complex astronomical image into a geometric problem in a high-dimensional feature space. This transformation provides a unified framework for addressing a wide range of astronomical challenges, from object detection and classification to anomaly identification in time series data.

\section{Discussion of Application Examples}

The versatility and effectiveness of this approach in various astronomical applications can be really large. In principle, it is suitable for any task in which there is the need to classify or extract information from an image. In this section, we discuss a few examples of applications that we tackled in the past few years.

\subsection{AGN Detection}

\begin{figure}
    \centering
    \includegraphics[width=.6\linewidth]{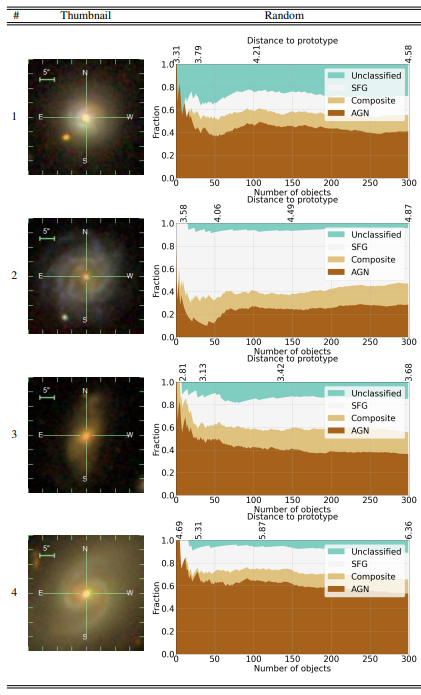}
    \caption{Some examples of the application of \texttt{ULISSE} to the detection of AGNs. See \cite{Doorenbos22}, from which the image is extracted, for further details.}
    \label{fig:ulissepaper1}
\end{figure}

In our initial study \cite{Doorenbos22}, we applied \texttt{ULISSE} to the detection of Active Galactic Nuclei (AGN) using multi-band images from the Sloan Digital Sky Survey (SDSS). The results demonstrated \texttt{ULISSE}'s capability to identify AGN candidates based on a single image. Despite not having been trained specifically for this, this is somehow related to a combination of several properties of the image host galaxy morphology, color (since we make use of three bands images, specifically \textit{g}, \textit{r} and \textit{i}), and the presence of a central nuclear source. The method showed a retrieval efficiency that was really competitive considering the usage of SDSS images in three bands, so a very limited amount of information for such a complex task as the AGN detection. 
The results of the method clearly depend on the morphology of the object, as can be seen from Fig.~\ref{fig:ulissepaper1}, being able to select candidates, with contribution from AGN emission, on average higher than $33\%$ for prototype objects with different morphology (and up to $60\%$ considering the composite objects selected by the BPT-diagram). On the other hands, the same application of \texttt{ULISSE} to the set of ``true'' non-AGN objects returned only $20\%$ of AGN candidates as false positives. It goes without saying that we choose to test the method considering sources that can be considered \textit{tricky}. See \cite{Doorenbos22} for further details.

\FloatBarrier 
\subsection{Estimation of local galaxy properties}

\begin{figure}[bh!]
    \centering
    \includegraphics[width=.47\textwidth]{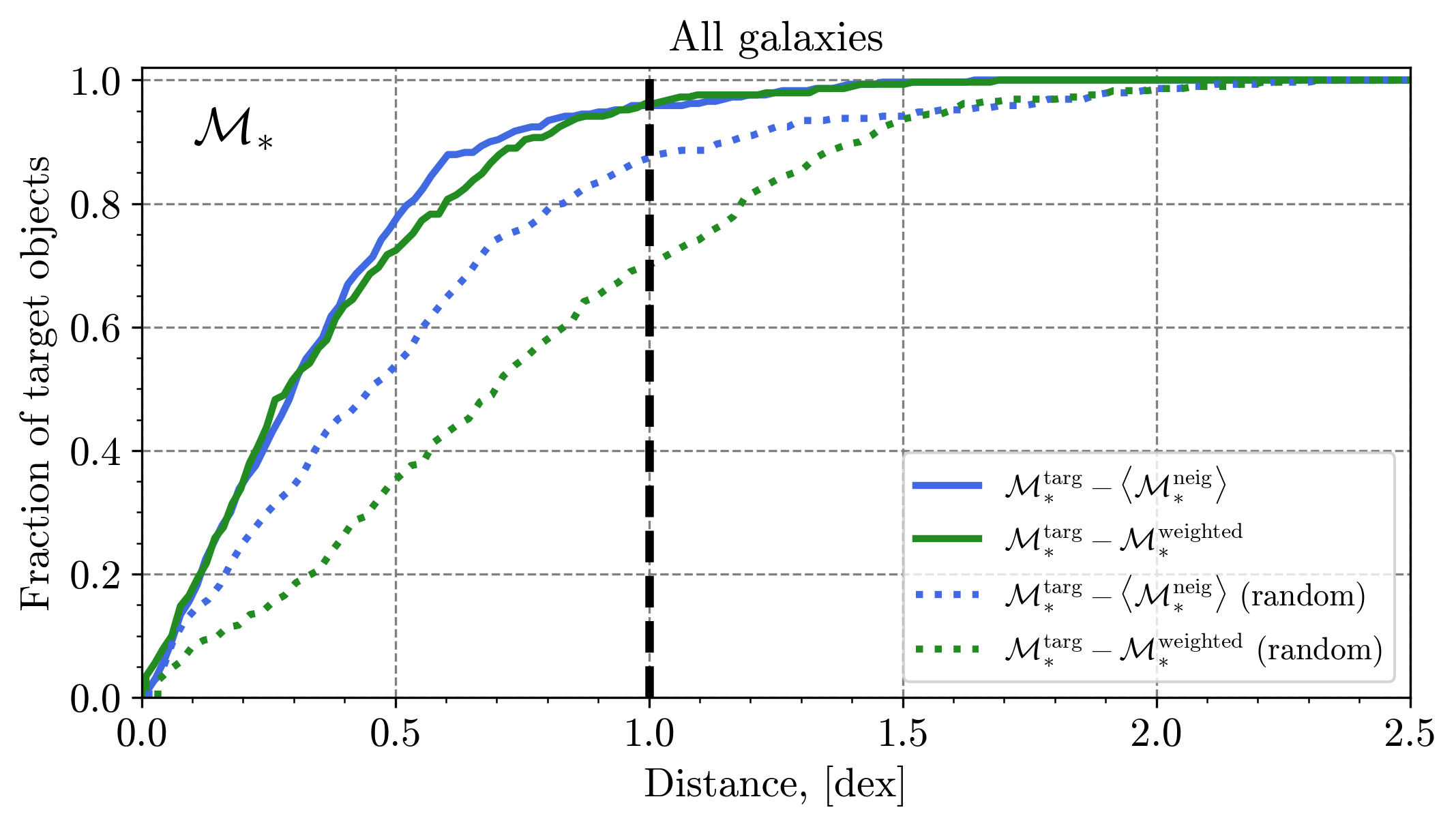}
    \includegraphics[width=.47\textwidth]{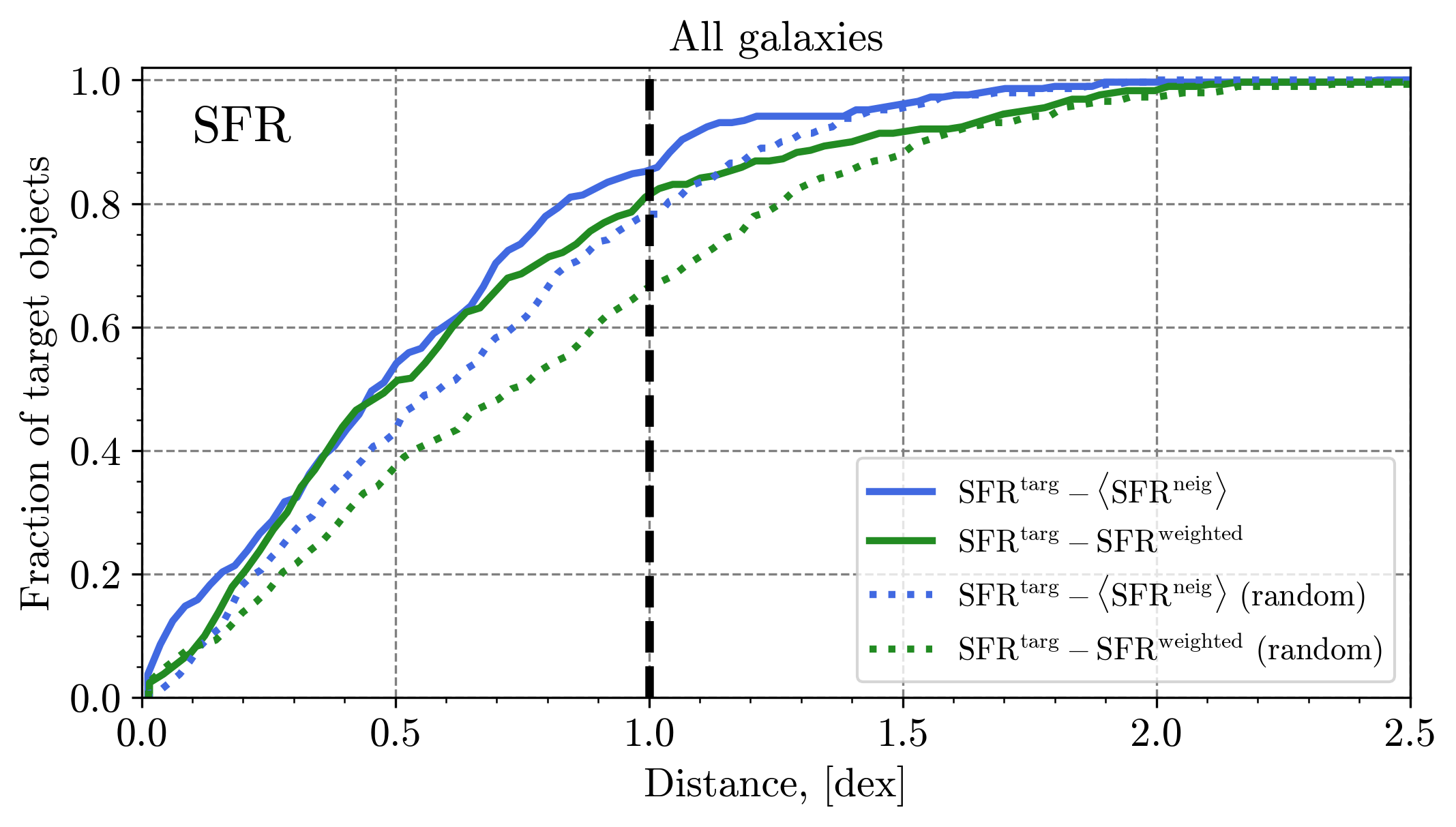}\\
    \caption{{\it Left panel:} The distributions of distances between the mean and the weighted mean $\mathcal{M}_{\ast}$ (weighted for the heterogeneous distribution of galaxies in the SFR-$\mathcal{M}_{\ast}$ parameter space in our primary sample) for the retrieved neighbors and the target object, i.e. $\mathcal{M}_{\ast}^{\rm targ} - \langle\mathcal{M}_{\ast}^{\rm neig}\rangle$ (by blue color) and $\mathcal{M}_{\ast}^{\rm targ} - \mathcal{M}_{\ast}^{\rm weighted}$ (by green color). {\it Right panel:} The same as on the left panel, but for ${\rm SFR}^{\rm targ} - \langle{\rm SFR}^{\rm neig}\rangle$ and ${\rm SFR}^{\rm targ} - {\rm SFR}^{\rm weighted}$. See \cite{torbaniuk2024} for further details.} \label{fig:hist-dist-M-SFR}
\end{figure}

Building upon the success in AGN detection, in a recent work \cite{torbaniuk2024} we extended \texttt{ULISSE}'s application to the estimation of galaxy properties. Our reference sample is based on the galSpec catalogue of galaxy properties\footnote{\url{https://www.sdss.org/dr12/spectro/galaxy_mpajhu/}} which was produced by the MPA-JHU group as the subsample from the main galaxy catalogue of the 8th Data Release of the Sloan Digital Sky Survey (SDSS DR8 \cite{Brinchmann2004}). This study demonstrates that the feature space created by \texttt{ULISSE} can be used to derive the stellar mass and the Star Formation Rate (SFR) for galaxies. This application highlights the richness of information captured in the \texttt{ULISSE}-generated feature space, extending beyond simple classification tasks to quantitative parameter estimation. Fig.~\ref{fig:hist-dist-M-SFR} summarizes the results in terms of the distribution of the distances between the estimate for the target object and the averaged one for the set of the retrieved neighbours showing it to be smaller than 1 dex for  more than $80\%$ of  the studied target objects , both in terms of SFR and stellar mass.

\subsection{Strong Gravitational Lensing Detection}

\begin{figure}
    \centering
    \includegraphics[trim={0 5.8cm 0 0}, clip, width=1\linewidth]{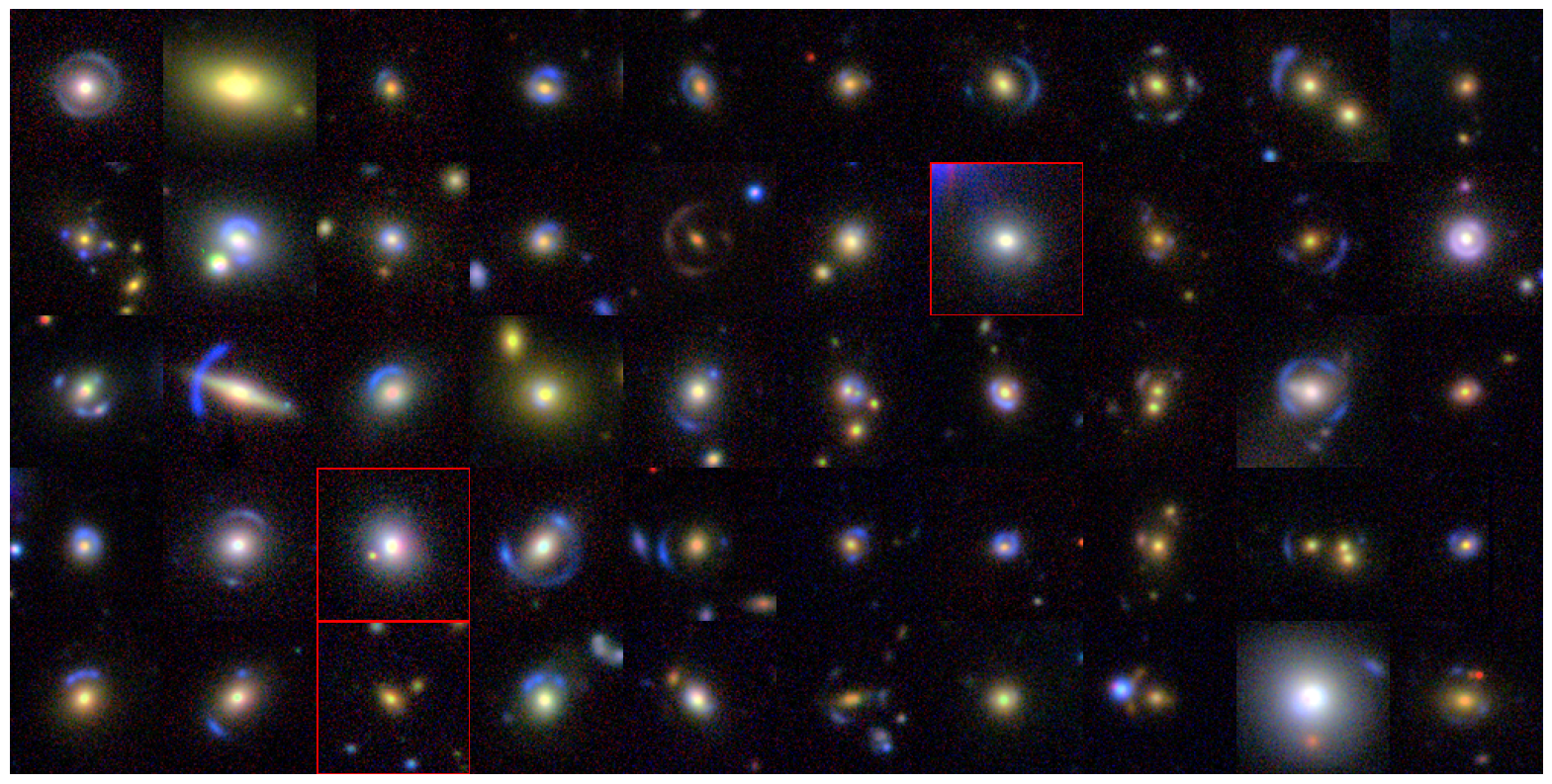}
    \caption{Top forthy images, in terms of frequency, proposed by \texttt{ULISSE}. The false positives have been framed within a red rectangle. }
    \label{fig:50-mostfrequent}
\end{figure}

We started the application of \texttt{ULISSE} in the search for strong-gravitational lensing events. Although in the work from Sasanelli et al. which is still in preparation, \texttt{ULISSE} has shown remarkable performance in identifying strong lensing candidates in simulated data. Concerning the results, within the first hundred most frequent candidates proposed by
\texttt{ULISSE}, on the positive queries, only six are negatives (i.e. $94\%$ accuracy); $14$ ($93\%$) in the first two-hundred. Fig.~\ref{fig:50-mostfrequent} shows the most $50$ frequent strong-lensing events proposed by \texttt{ULISSE}.

However, the transition to real observational data has presented additional challenges, likely due to the small amount of real lenses to be identified as \textit{sosia} of the starting one. The ongoing work is focused on refining the approach to bridge the gap between simulated and real-world performance.

\FloatBarrier 

\subsection{Anomaly Detection in Time Series}

One other application of transfer learning in astronomy is in the analysis of time series data. Astronomical time series, which represent the variation of physical quantities over time for celestial objects, are fundamental to understanding the dynamics of various astrophysical sources, from stars and planets to supernovae and galaxies \cite{Scargle97, Aigrain23}. However, these time series often suffer from artifacts caused by instrumental errors, atmospheric conditions, or contaminant sources, which can lead to false or misleading conclusions if not properly addressed. 

In the paper \cite{cavuoti2024}, we made use of the same feature extractor in order to identify anomalies in data from VLT Survey Telescope (VST) that, with its wide-field imaging capability, produces time series data useful for studying variable stars and transient events \cite{CapaccioliSchipani15, decicco2021}.


By transforming each epoch of a light curve into the feature space, we can identify outlier points that deviate significantly from the typical behavior. We defined as the ``typical behaviour'' the stacked image that hence defines a dynamical threshold that filters out the anomalies. This approach has shown great potential in flagging problematic observations, such as those affected by instrumental errors or transient phenomena, without the need for manual inspection at the price of very few good epochs lost. This capability is particularly valuable for large-scale surveys where manual inspection of every light curve is, of course, infeasible.

\subsection{Large-scale Anomaly Detection}

\begin{figure}
    \centering
    \includegraphics[width=1\linewidth]{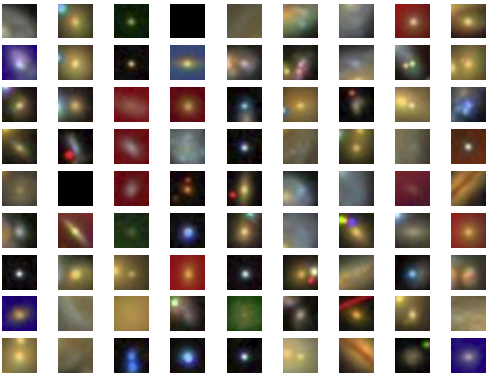}
    \caption{An example of the top $81$ most \textit{weird} images in a dataset extracted from SDSS, as it can be easily seen most of them contain some obvious problems}
    \label{fig:anomaly-detection}
\end{figure}

Looking to the future, one of the most promising potential applications of \texttt{ULISSE} is in the detection of anomalies in extremely large astronomical datasets. As we move into the era of big data astronomy with projects like the Rubin Observatory Legacy Survey of Space and Time (LSST), the ability to efficiently search for rare or unknown phenomena becomes increasingly important. \texttt{ULISSE}'s approach of transforming complex astronomical data into a well-structured feature space is well-suited for this task. By defining a sort \textit{normality} in this feature space, as for instance a low distance with respect to the first neighbour, we can potentially identify objects or events that deviate significantly from expected patterns, possibly leading to the discovery of new classes of astronomical phenomena or rare events. We tested, for a preliminary proof of concept, this approach in SDSS with the result of identifying a lot of problematic images as can be seen from Fig.~\ref{fig:anomaly-detection}; a recursive work should be able to filter out a lot of problematic images (that is, anyway, a useful by-product) and then identify rare or even unknown kind of objects.

\subsection{Future Prospects}

The examples discussed here demonstrate the broad applicability of a transfer learning approach across various domains of astronomical research. This approach provides a flexible framework for extracting meaningful information from complex astronomical data, from object detection and classification to parameter estimation and anomaly detection.

As we continue to refine and expand the capabilities of this kind of approach, several areas of future development are particularly promising: \textbf{Multi-wavelength Integration:} Extending \texttt{ULISSE} to directly incorporate data from multiple wavelengths could provide even richer feature representations, the current version handles up to three bands. \textbf{Time Domain Astronomy:} Further development of this approach's capabilities in analyzing time series data could improve our approach to variability studies and transient classification. \textbf{Improve the feature extractor}: despite the results that we achieved with \texttt{ImageNet} being really good, there are new models being produced that could be worth to be used.

In conclusion, this approach represents a powerful and versatile tool for astronomical data analysis. Its ability to transform complex astronomical data into a rich but easily manageable feature space opens up new possibilities for discovery and analysis across a wide range of astronomical applications. As we continue to explore and refine this approach, we think that it has the potential to play a significant role in unlocking the full scientific potential of current and future astronomical datasets.

\bibliographystyle{ws-procs961x669}

\bibliography{cavuoti} 

\begin{thebibliography}{10}

\bibitem{York2000}
D.~G. {York}, J.~{Adelman}, J.~E. {Anderson}, Jr. {\em et~al.}, {The Sloan Digital Sky Survey: Technical Summary}, {\em \aj} {\bf 120}, 1579 (September 2000).

\bibitem{deJong2015}
J.~T.~A. {de Jong}, G.~A. {Verdoes Kleijn}, D.~R. {Boxhoorn} {\em et~al.}, {The first and second data releases of the Kilo-Degree Survey}, {\em \aap} {\bf 582}, p. A62 (October 2015).

\bibitem{Aihara2019}
H.~{Aihara}, Y.~{AlSayyad}, M.~{Ando} {\em et~al.}, {Second data release of the Hyper Suprime-Cam Subaru Strategic Program}, {\em \pasj} {\bf 71}, p. 114 (December 2019).

\bibitem{Ivezic2019}
{\v{Z}}.~{Ivezi{\'c}}, S.~M. {Kahn}, J.~A. {Tyson} {\em et~al.}, {LSST: From Science Drivers to Reference Design and Anticipated Data Products}, {\em \apj} {\bf 873}, p. 111 (March 2019).

\bibitem{scaramella2022euclid}
R.~Scaramella, J.~Amiaux, Y.~Mellier {\em et~al.}, Euclid preparation-i. the euclid wide survey, {\em Astronomy \& Astrophysics} {\bf 662}, p. A112  (2022).

\bibitem{AlvarezMarquez2019}
J.~{{\'A}lvarez-M{\'a}rquez}, L.~{Colina}, R.~{Marques-Chaves} {\em et~al.}, {Investigating the physical properties of galaxies in the Epoch of Reionization with MIRI/JWST spectroscopy}, {\em \aap} {\bf 629}, p.~A9 (September 2019).

\bibitem{Baron2019}
D.~{Baron}, {Machine Learning in Astronomy: a practical overview}, {\em arXiv e-prints} , p. arXiv:1904.07248 (April 2019).

\bibitem{Longo2019}
G.~{Longo}, E.~{Mer{\'e}nyi} and P.~{Ti{\v{n}}o}, {Foreword to the Focus Issue on Machine Intelligence in Astronomy and Astrophysics}, {\em \pasp} {\bf 131}, p. 100101 (November 2019).

\bibitem{Fluke2020}
C.~J. {Fluke} and C.~{Jacobs}, {Surveying the reach and maturity of machine learning and artificial intelligence in astronomy}, {\em WIREs Data Mining and Knowledge Discovery} {\bf 10}, p. e1349 (January 2020).

\bibitem{Doorenbos22}
L.~{Doorenbos}, O.~{Torbaniuk}, S.~{Cavuoti} {\em et~al.}, {ULISSE: A tool for one-shot sky exploration and its application for detection of active galactic nuclei}, {\em \aap} {\bf 666}, p. A171 (October 2022).

\bibitem{tan2019efficientnet}
PMLR, {\em Efficientnet: Rethinking model scaling for convolutional neural networks} 2019.

\bibitem{Schmidhuber2015}
J.~Schmidhuber, Deep learning in neural networks: An overview, {\em Neural Networks} {\bf 61}, 85  (2015).

\bibitem{McCulloch1943}
W.~S. McCulloch and W.~Pitts, A logical calculus of the ideas immanent in nervous activity, {\em Bulletin of Mathematical Biophysics} {\bf 5}, 115  (1943).

\bibitem{Rosenblatt1958}
F.~Rosenblatt, The perceptron: A probabilistic model for information storage and organization in the brain, {\em Psychological Review} , 65  (1958).

\bibitem{deng2009imagenet}
Ieee, {\em Imagenet: A large-scale hierarchical image database} 2009.

\bibitem{torbaniuk2024}
O.~Torbaniuk, L.~Doorenbos, M.~Paolillo {\em et~al.}, Ulisse: Determination of star-formation rate and stellar mass based on the one-shot galaxy imaging technique, {\em \aap}   (2024 submitted).

\bibitem{Brinchmann2004}
J.~Brinchmann, S.~Charlot, S.~White {\em et~al.}, The physical properties of star-forming galaxies in the low-redshift universe, {\em Monthly Notices of the Royal Astronomical Society} {\bf 351}, 1151  (2004).

\bibitem{Scargle97}
J.~D. Scargle, Astronomical time series analysis, in {\em Astronomical Time Series\/},  eds. D.~Maoz, A.~Sternberg and E.~M. Leibowitz (Springer Netherlands, Dordrecht, 1997).

\bibitem{Aigrain23}
S.~{Aigrain} and D.~{Foreman-Mackey}, {Gaussian Process Regression for Astronomical Time Series}, {\em \araa} {\bf 61}, 329 (August 2023).

\bibitem{cavuoti2024}
S.~{Cavuoti}, D.~{De Cicco}, L.~{Doorenbos} {\em et~al.}, {Identification of problematic epochs in astronomical time series through transfer learning}, {\em \aap} {\bf 687}, p. A246 (July 2024).

\bibitem{CapaccioliSchipani15}
M.~{Capaccioli} and P.~{Schipani}, {The VLT Survey Telescope Opens to the Sky: History of a Commissioning}, {\em The Messenger} {\bf 146}, 2 (December 2011).

\bibitem{decicco2021}
D.~De~Cicco, F.~Bauer, M.~Paolillo {\em et~al.}, A random forest-based selection of optically variable agn in the vst-cosmos field, {\em Astronomy and Astrophysics} {\bf 645}  (2021).

\end{thebibliography}

\vfill
\pagebreak


\end{document}